\documentclass[12pt]{iopart}
\usepackage{epsf}
\begin{document}

\newcommand{\ua}{\uparrow}
\newcommand{\da}{\downarrow}
\newcommand{\la}{\langle}
\newcommand{\ra}{\rangle}
\newcommand{\rar}{\rightarrow}
\newcommand{\us}{\uparrow}
\newcommand{\ds}{\downarrow}
\newcommand{\be}{\begin{equation}}
\newcommand{\ee}{\end{equation}}
\newcommand{\bea}{\begin{eqnarray}}
\newcommand{\eea}{\end{eqnarray}}

\title{Coexistence of 
antiferromagnetism and superconductivity
in the Anderson lattice}

\author{P. D. Sacramento}
\address{Departamento de F\'{\i}sica and  CFIF, Instituto Superior 
T\'ecnico, Av. Rovisco Pais, 1049-001 Lisboa, Portugal}

\date{\today}


\begin{abstract}
We study the interplay between antiferromagnetism and superconductivity
in a generalized infinite-$U$ Anderson lattice, where both 
superconductivity and  antiferromagnetic order are introduced phenomenologically
in mean field theory. In a certain regime, a quantum phase transition is found 
which is characterized by an abrupt expulsion of magnetic order by  $d$-wave 
superconductivity, 
as externally applied pressure increases. This transition 
takes place when the $d$-wave superconducting critical temperature, 
$T_c$, intercepts the magnetic critical temperature, $T_m$, under increasing pressure. 
Calculations of the  quasiparticle
bands and   density of states in the ordered phases are presented.
We calculate the optical conductivity $\sigma(\omega)$
in the clean limit. It is shown that when the temperature drops below
$T_m$ a double peak structure develops in $\sigma(\omega)$.
\end{abstract}
\vspace{0.3cm}
\section{Introduction}

It is a common feature of several strongly correlated electronic systems
that the low temperature ordered phases compete with each other. In particular,
competition between antiferromagnetic and superconducting orders is an
important characteristic of  heavy-fermion systems\cite{Lonzarich}, 
which is also shared by high-$T_c$ materials \cite{hightc}
 and low-dimensional systems \cite{Jerome}. 
The closeness of the superconducting phase to the antiferromagnetic phase 
in heavy fermion compounds has lead
to the  conjecture that the attractive interaction leading to superconductivity
is actually mediated by a magnetic excitation \cite{Lonzarich}, instead of the 
traditional phonon mechanism. Also, the interplay between antiferromagnetism
and superconductivity has been proposed to be described by a $SO(5)$ symmetry
breaking \cite{Zhang}. The complexity of these
systems arises from the interplay between Kondo screening of the local moments,
the antiferromagnetic (RKKY) interaction between the local moments and 
superconducting correlations between the heavy quasiparticles.

Heavy fermion systems that exhibit both superconductivity  and  antiferromagnetism 
exhibit ratios between the N\'eel temperature $T_N$
and the superconducting critical temperature $T_c$  that can vary substantially
(of the order of $T_N/T_c\sim 1-100$),
 with  coexistence of both types of order below $T_c$.
The coexistence of both types of order can be tuned  by external
parameters such as externally applied pressure or chemical pressure  
(involving changes in the stoichiometry).\cite{Lonzarich,ishida99}
Examples of  heavy-fermion materials 
which exhibit antiferromagnetic and superconducting order 
at low temperature  are  $URu_2Si_2$ and  $U_{0.97}Th_{0.03}Be_{13}$.
It has recently been found that $UPd_2Al_3$   
($T_N=14.3$ K and $T_c=2$ K) and $UNi_2Al_3$  ($T_N=4.5$ K and $T_c=1.2$ K)
show coexistence of superconductivity and local moment antiferromagnetism. 
\cite{Lonzarich,steglich93,feyerherm94,bernhoeft98,also00} 
However, in the $Ce$-based heavy-fermions magnetism tipically competes
with superconductivity.
In the prototype heavy-fermion system Ce$_x$Cu$_2$Si$_2$ both coexistence
and  competition between $d-$wave superconductivity  and magnetic order 
has been clearly observed in a small range of $x$ values
around $x\simeq 0.99$  pressure.\cite{ishida99} 
This system exhibits a magnetic ``A phase'' at low temperature
whose detailed nature is not yet known. Increasing pressure reduces
the critical temperature $T_A$ of the A phase.
Recent studies\cite{ishida99,gegenwart,bruls,luke}
of  $\rm CeCu_2Si_2$  samples near stoichiometric composition
have shown that a $d$-wave superconducting phase expels  the magnetic
``A phase'' when $T_A$ approaches $T_c$ under increasing pressure.

In general terms, the local moments due to the $f$-electrons
are progressively quenched as the temperature lowers. In
dilute systems the picture is well understood as due to the
Kondo screening by the conduction electrons. In dense
systems however the picture is more involved. At low
temperatures the local moments are not completely quenched.
In  $U$-based materials such as $URu_2Si_2$, $UPt_3$ or
$UPd_3$, for instance,  the remaining moments are quite small of the order of
$0.01-0.03 \mu_B$ but for other systems such as $UPd_2Al_3$ the
local moment is quite large of the order of $0.85 \mu_B$. 
The low temperature magnetic behaviour of $Ce$-based compounds 
such  as $CeCu_2Si_2$ and $CeCu_{6-x}Au_x$
has been interpreted as due to the vicinity
to a quantum critical point \cite{SteglichJ} where the N\'eel temperature
tends to zero. Two pictures
arise however \cite{ColemanCe}: in the first one the Kondo
temperature is high (the moments are quenched at a finite temperature)
and when the system approaches the quantum critical point there are
no free moments (assuming that quenching is complete). Then the
system has to order due to a Fermi surface instability of the spin density
wave type. Another possible situation is one in which the moments are
not completely quenched down to $T=0$ and are free to orient themselves
leading to magnetism. In the case of $CeCu_{6-x}Au_x$ evidence has been
recently found that the second picture seems to hold \cite{ColemanCe} but a
small but finite Kondo temperature has been quoted for this material
(see the scond reference in \cite{SteglichJ}). On the
other hand the high value of the Kondo temperature for the $CeCu_2Si_2$
compound \cite{ishida99} indicates possibly that the first scenario
should hold.  Furthermore,
recent experiments \cite{Pagliuso} with $CeRh_{1-x}Ir_xIn_5$
also reveal unusual coexistence of magnetism and superconductivity.
It appears that in this system the $f$-electrons are more
band-like than localized. On the other hand in $UPd_2Al_3$ it is the
dual character of the $5f$ electrons that leads to the high value of
the local moment in coexistence with the itinerant electrons and
with the superconductivity \cite{feyerherm94}. 

On the theoretical side it is believed that the Anderson model and its
extension to the lattice captures the basic physics of the heavy-fermions
\cite{NewnsRead}. In this model the conduction electrons, $c$, (usually
regarded as free) hybridize with local states, $f$, where the electrons are
strongly interacting in such a way that the Coulomb repulsion, $U$, between
two $f$-electrons is the largest energy scale in the problem. Frequently
the limit $U \rightarrow \infty$ is taken implying that double occupancy is
forbidden. The limit $U=\infty$ has been studied using the slave boson
technique.\cite{Millis87,coleman84,houghton88}
In particular, it has been 
shown that superconducting instabilities arise in the $p$ 
and $d$-wave channels because of the effective (RKKY) interaction between the 
$f-$electrons.\cite{houghton88,Lavagna}
Recently, the magnetic and superconducting
instabilities of the normal phase were studied in the 
random phase approximation (RPA) \cite{RPA}  
by taking into account slave boson fluctuations
above the condensate.  However,
the competition/coexistence between both types of ordering was not considered. 

In this work we consider the $U=\infty$ Anderson lattice model in the 
slave-boson approach. Because our aim is to 
study the  interplay between
magnetism and superconductivity, we explicitly introduce   antiferromagnetic  and 
superconducting couplings phenomenologically. 
The coupling constants are taken as independent
even though they are related if the superconducting mechanism is mediated via
the RKKY interaction. Using a mean-field approach we study directly the
ordered phases and determine regimes of coexistence or  competition between
the two ordered phases depending on the parameters of the model.

In this work we will take the conduction electrons to be non-interacting but
we should also mention that attempts to include
interactions between the conduction electrons have been carried out \cite{coulomb,oliveira}.
The inclusion of this more realistic interaction has been found to be required
in some systems to attain a better understanding of the experimental results.
We focus our attention on the coexistence of superconducting correlations and
magnetic ordering in heavy fermions which, to our knowledge,
has not yet been studied theoretically despite the considerable recent
experimental effort devoted to this subject.

\section{Model Hamiltonian and quasiparticle spectrum}

The microscopic description of superconductivity and magnetic order in
the  Anderson lattice model is a still unsolved problem. In the folllowing we
shall  consider  an effective
Hamiltonian which originates from  the $U=\infty$ Anderson model with two 
additional phenomenological terms: 
one where superconducting correlations are explicitly 
assumed between the local $f$-electrons (since it is believed that 
pairing occurs between heavy quasi-particles which, therefore, have essentially
$f$ character) and another term
where local spins are coupled antiferromagnetically. 
The infinite Coulomb repulsion between the $f$ electrons 
is described within Coleman's  slave-boson\cite{coleman84} technique with
a condensation amplitude $\sqrt{z}$.
The  effective Hamiltonian is therefore:
\bea
H^{MF} & = & \sum_{k\sigma} (\epsilon_k - \mu) c_{k \sigma}^{\dagger} c_{k\sigma}
+ \sum_{k\sigma} (\epsilon_f - \mu) f_{k \sigma}^{\dagger} f_{k\sigma} \nonumber \\
& + & \sqrt{z} V \sum_{k\sigma} \left( f_{k \sigma}^{\dagger} c_{k\sigma}
+ c_{k \sigma}^{\dagger} f_{k\sigma} \right) \nonumber \\
& + & z \sum_{k} \left( \Delta_f \eta_k f_{k \uparrow}^{\dagger} 
f_{-k,\downarrow}^{\dagger}
+ \Delta_f^{*} \eta_k f_{-k,\downarrow} f_{k \uparrow} \right) \nonumber \\
& - & 2 m \sum_{k \sigma} \sigma \left( f_{k \sigma}^{\dagger} f_{k-Q,\sigma} +
f_{k \sigma}^{\dagger} f_{k+Q,\sigma} \right) \nonumber \\
& + & (\epsilon_f-\epsilon_0)(z-1)N_s + \frac{2 N_s m^2}{J_m} 
 -  \frac{N_s |\Delta_f|^2}{J_f} \label{model}
\eea
The $c$ and $f$ operators refer to conduction and localized electrons and obey the 
usual anticommutation 
relations. For simplicity, the hybridization potential $V$ is assumed to be momentum
independent, $\epsilon_0$ and  $\epsilon_f$ denote the bare and renormalized 
 $f$-level  energies, and $N_s$ denotes the number of lattice sites.

Although it is known that conduction electrons provide an effective RKKY 
interaction between $f$ electrons, 
we remark that the superconducting and magnetic order parameters  in (\ref{model})
{\it cannot} be {\it simultaneously}
derived from a Hubbard-Stratonovich decoupling of a single
RKKY term of the form ${\tilde J}_{ij} \vec{S}_i \cdot \vec{S}_j$, 
as discussed, for instance, in ref. \cite{Ubbens} This is why we have 
phenomenologically introduced those terms.

In writing down the pairing term in (\ref{model}) we have in mind that 
in the slave boson formulation a slave-boson operator  
$b_i$ is  associated with every $f_i$  operator to prevent double occupancy.
Condensation of the slave bosons is described  by the replacement 
$b_i \rightarrow <b_i>=<b_i^{\dagger}>= \sqrt{z}$, hence the factor $z$ in 
the superconducting term of (\ref{model}).\cite{peres00,Ruckenstein}
The superconducting order
parameter is given by $\Delta_f=\frac{zJ_f}{N_s} \sum_k \eta_k <f_{-k,\downarrow}
f_{k\uparrow}>$, where $\eta_k$ denotes any of the possible pairing symmetries
 $\eta_k^{(s)}=\cos{k_x}+\cos{k_y}$, $\eta_k^{(p,i)}=\sqrt{2}
\sin{k_i}$ and $\eta_k^{(d)}=\cos{k_x}-\cos{k_y}$ for $s$, $p$ and $d$ waves,
respectively. Here we consider two dimensions for simplicitly of the calculations.
We take a square lattice even though several heavy-fermions have a complicated
lattice structure since we want to capture the main features.
 
The magnetic order parameter is given by the  mean-field equation
$m_i = J_m < f_{i \uparrow}^{\dagger} f_{i\uparrow}-
f_{i \downarrow}^{\dagger} f_{i\downarrow}>$
and  $m_i=2 m \cos(\vec{r}_i \cdot \vec{Q})$ where $\vec{Q}=(\pi,\pi)$ 
is the antiferromagnetic ordering vector. We consider only commensurate
antiferromagnetic correlations since it is particularly relevant to the
systems referred.

In the calculations we assume a
simple dispersion for the conduction electrons, of the
form $\epsilon_k=-2t \sum_{i=x,y} \cos{k_i}$.
For a given particle density the chemical potential must be computed
from the condition $n_c+n_f=n$.
The mean-field equations obtained from minimization of the free energy 
of Hamiltonian (\ref{model})   are solved numerically. 
 The solution gives the  interplay between the boson condensation, 
the magnetization, and  superconducting pairing as  function of band-filling
and of the various model  parameters. Throughout this work we will
consider only the case of $d$-wave pairing. 
The other two pairings give qualitatively similar results in most regimes
as stressed before for the case of no magnetism and only superconducting order
\cite{peres00}. In different regimes different pairing symmetries become
the most stable one but the overall behavior is similar. However we will get
back to this point later.
We will consider temperatures
such that the slave bosons are condensed and therefore we quench the
$f$-electron density fluctuations. 

\begin{figure}
\epsfxsize=8.0cm
\epsfysize=8.0cm
\centerline{\epsffile{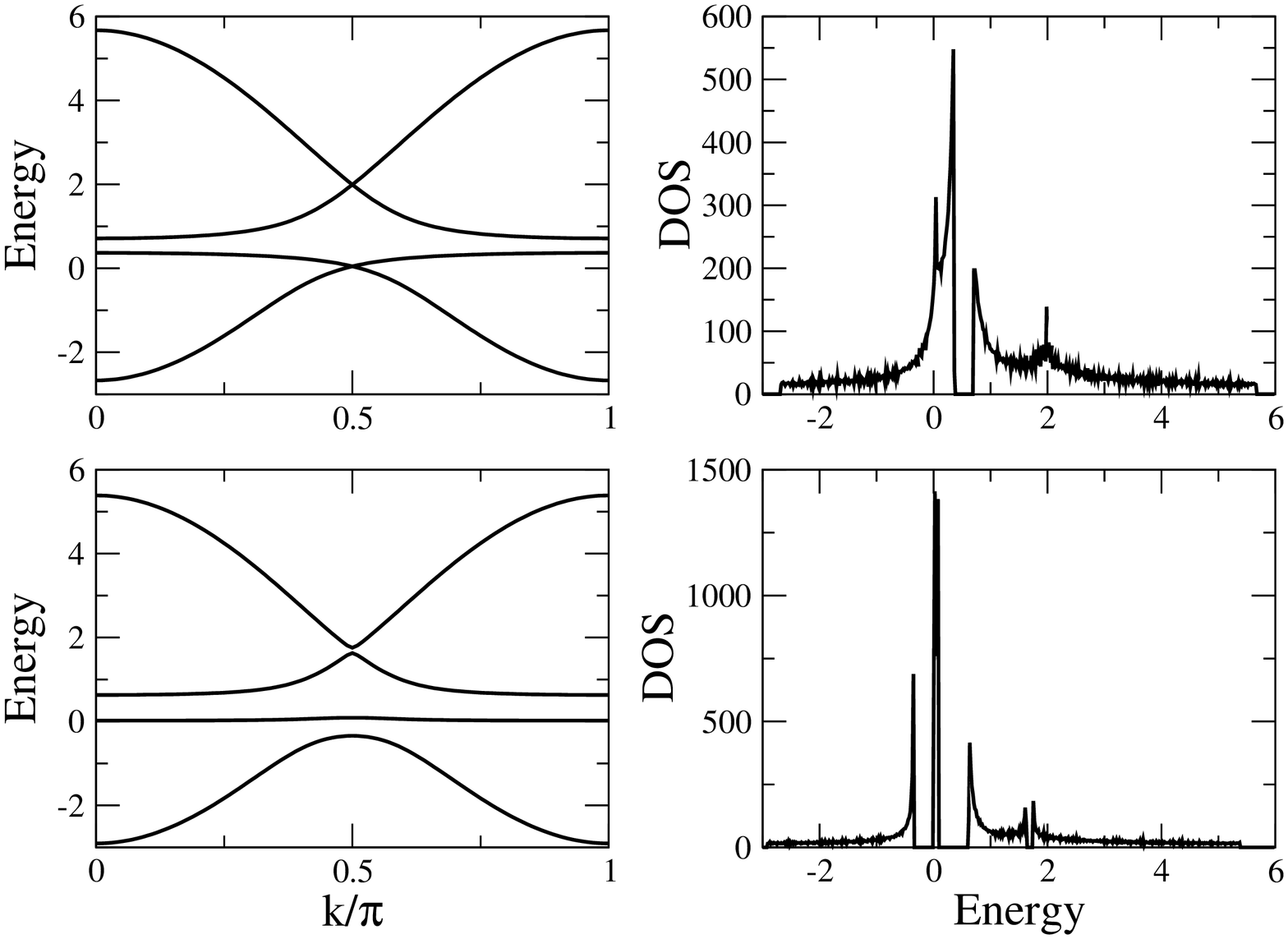}}
\caption{ Quasiparticle bands and density of states for the magnetic 
coupling $J_m=0$ in the top panels and $J_m=0.9$ in the lower panels 
along the direction $k_x=k_y=k$. Model parameters are: 
$t=1$, $n=0.9$, $V=1.0$, and $\epsilon_0=-1.5$. The energies
are measured with respect to the chemical potential.
}
\label{jm}
\end{figure}

\subsection{Magnetic non-superconducting phase}

 We  begin  by discussing the magnetic but non-superconducting phase
of the model. A detailed study of the superconducting non-magnetic phases has
been presented  elsewhere.\cite{oliveira,peres00}

In order to diagonalize  (\ref{model}) for $J_f=0$ we introduce a quasi-particle 
operator which is a linear combination of the 
operators forming  the basis 
$\left(c_{k\sigma}^{\dagger}; c_{k+Q,\sigma}^{\dagger};
f_{k\sigma}^{\dagger}; f_{k+Q,\sigma}^{\dagger} \right)$.
We write the Hamiltonian in this basis.
The mean-field equations are obtained 
varying the effective Hamiltonian in the usual way. Minimization of the free energy
of (\ref{model}) with respect to $z$ gives:
\begin{equation}
\epsilon_f-\epsilon_0 = - \frac{V}{2\sqrt{z}} \frac{1}{N_s}
\sum_{k\sigma} \left( <f_{k\sigma}^{\dagger}c_{k\sigma}> +
<c_{k\sigma}^{\dagger}f_{k\sigma}> \right) \,.
\label{varz}
\end{equation}
The condition $n_f+z=1$ reads:
\begin{equation}
z=1-\frac{1}{N_s} \sum_{k\sigma} <f_{k\sigma}^{\dagger} f_{k\sigma}> \,,
\end{equation}
and the chemical potential is related to  the total particle density as:  
\begin{equation}
n=\frac{1}{N_s}\sum_{k\sigma} <c_{k\sigma}^{\dagger}c_{k\sigma}> 
+ \frac{1}{N_s} \sum_{k\sigma} <f_{k\sigma}^{\dagger}f_{k\sigma}> \nonumber\,.
\end{equation}
The magnetization is given by:
\begin{equation}
m=\frac{J_m}{2N_s} \sum_{k \sigma} 
\sigma < f_{k \sigma}^{\dagger} f_{k-Q,\sigma} +
f_{k \sigma}^{\dagger} f_{k+Q,\sigma} >\,.
\end{equation}

For the sake of comparison, we show in Figure \ref{jm} the
quasiparticle bands and the density of states for a
situation where the magnetic order parameter is zero 
and  nonzero, respectively. 
We see that in addition to the hybridization gap, there is an additional
gap  due to the magnetic order. The magnetic order with momentum
$\vec{Q}$ reduces the size of the  Brillouin zone to a 
half of that of a non-magnetic
system.

\begin{figure}
\epsfxsize=8.0cm
\epsfysize=8.0cm
\centerline{\epsffile{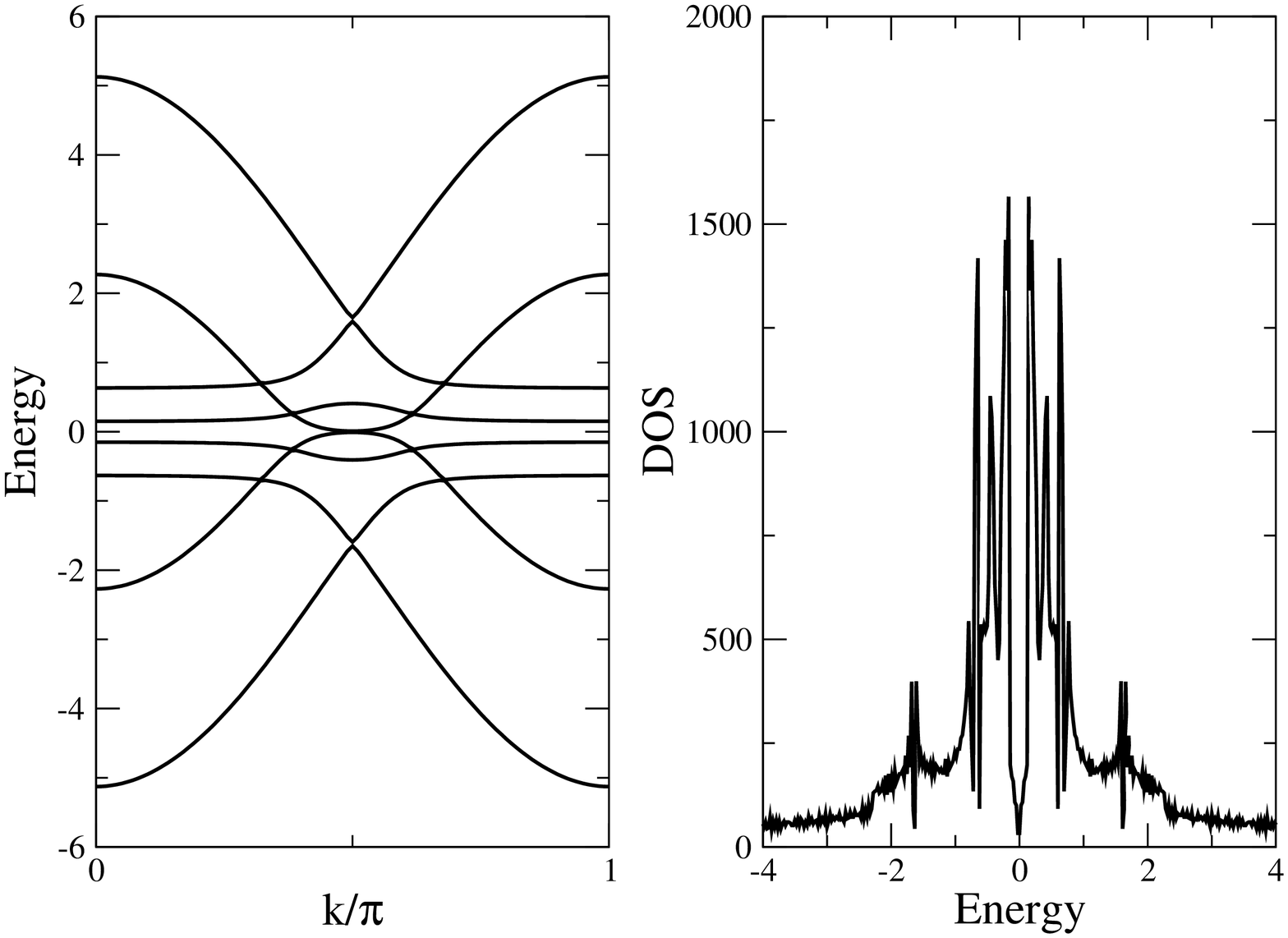}}
\caption{Quasiparticle bands and density of states for the
couplings $J_m=0.9$, $J_f=-3$ for the Anderson
model with magnetism and superconductivity. The other parameters are 
$n=0.9$, $V=0.6$,
$\epsilon_0=-1.5$ and the hopping $t=V/0.66$. 
}
\label{scmagbands}
\end{figure}

\subsection{Coexistence of antiferromagnetism and superconductivity}

The Hamiltonian matrix
 $H_k$ can be written  in the basis 
\[ \left(
c_{k\uparrow}, 
c_{k+Q,\uparrow},
f_{k,\uparrow}, 
f_{k+Q,\uparrow},
c_{-k,\downarrow}^{\dagger},
c_{-k-Q,\downarrow}^{\dagger},
f_{-k,\downarrow}^{\dagger},
f_{-k-Q,\downarrow}^{\dagger}\right)  \]
as:
\[ H_{k} = \left( \begin{array}{cc}
A_+ & D \\
D^{\dagger} & A_- \\
\end{array} \right)
\]
where the matrices $A_{\pm}$ and $D$ are given by
\[ A_{\pm} = \left( \begin{array}{cccc}
\pm \epsilon_k & 0 & \pm \sqrt{z}V & 0 \\
0 & \pm \epsilon_{k+Q} & 0 & \pm \sqrt{z}V  \\
\pm \sqrt{z}V & 0 & \pm \epsilon_f & -2m \\ 
0 & \pm \sqrt{z}V & -2m & \pm \epsilon_f \\
\end{array} \right) 
\]
and
\[ D = \left( \begin{array}{cccc}
0 & 0 & 0 & 0 \\
0 & 0 & 0 & 0 \\
0 & 0 & z\Delta_f \eta_k & 0 \\ 
0 & 0 & 0 & z \Delta_f \eta_{k+Q} \\ 
\end{array} \right) 
\]
The energies are measured with respect to the chemical potential. 
The eigenvectors of the matrix $H_k$  are the Bogolubov  operators  
expressed in the same basis.

Variation of the free energy with respect to $z$ shows that 
(\ref{varz}) must now  be replaced by
\begin{eqnarray}
\epsilon_f-\epsilon_0 &=&-\frac V{2\sqrt{z}N_s}
\sum_{\vec k,\sigma}\left(<f_{\vec k,\sigma}^{\dag}c_{\vec k,\sigma}>+
<c_{\vec k,\sigma}^{\dag}f_{\vec k,\sigma}>\right) \nonumber\\
&-& \frac{2N_s}{zJ_f}   \Delta_f^2
\end{eqnarray}
while the other mean-field equations remain unaltered.

In Figure \ref{scmagbands}
we show a typical quasiparticle band structure  for the  case where there
is coexistence of magnetism and superconductivity. 
The quasiparticle bands are symmetric around the chemical
potential due to the particle-hole structure of the Bogoliubov operators.
Due to the superconducting order the spectrum is gapless.

\section{Phase diagrams}
 
\begin{figure}
\epsfxsize=8.0cm
\epsfysize=8.0cm
\centerline{\epsffile{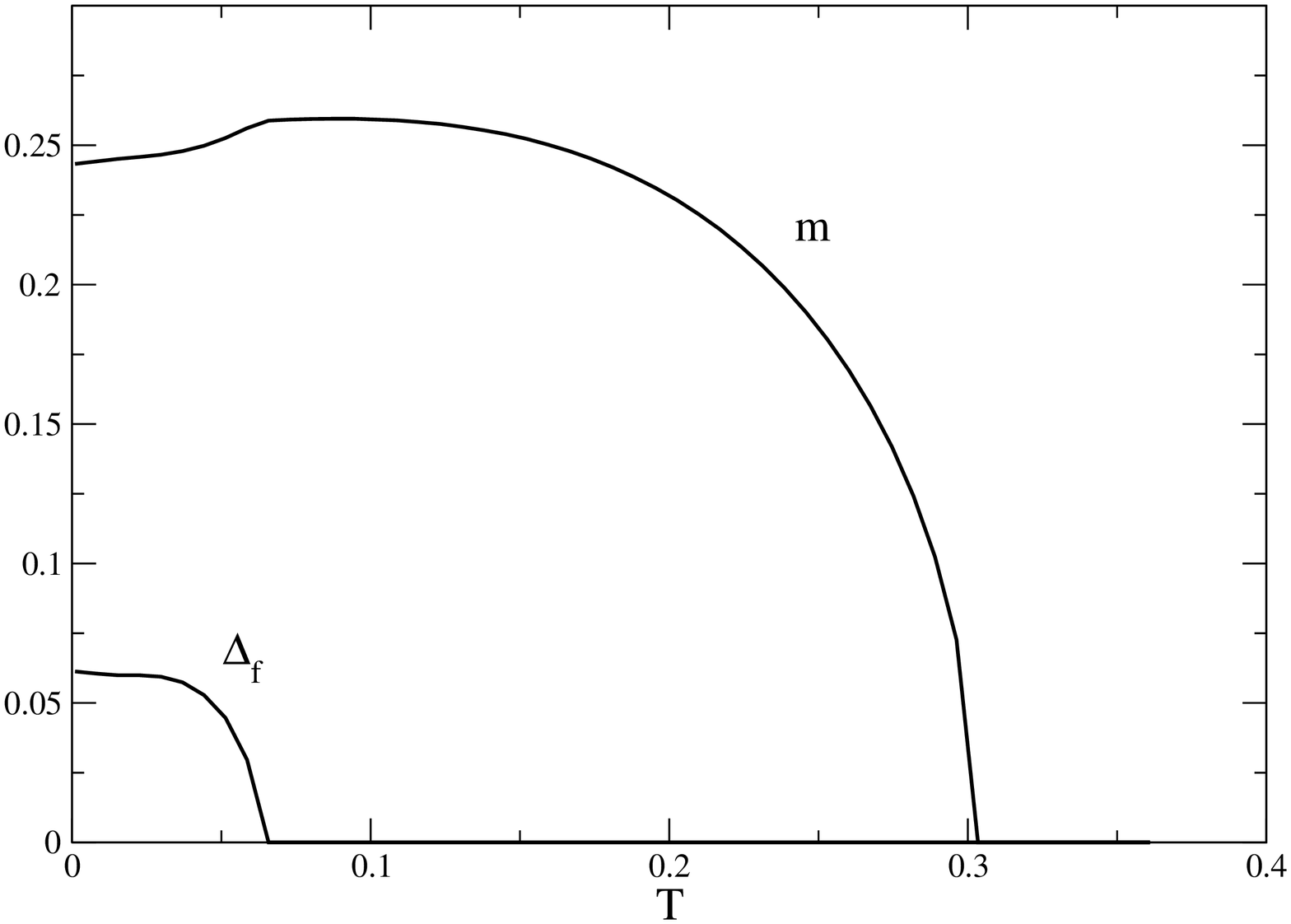}}
\caption{Magnetization and $d-$wave superconducting order parameters as 
functions
of temperature for $n=1$, 
$t=1$, $\epsilon_0=-1.5$, $J_f=-3$, $J_m=0.9$,
$V=0.66$.
}
\label{mDelta}
\end{figure}

\begin{figure}
\epsfxsize=8.0cm
\epsfysize=8.0cm
\centerline{\epsffile{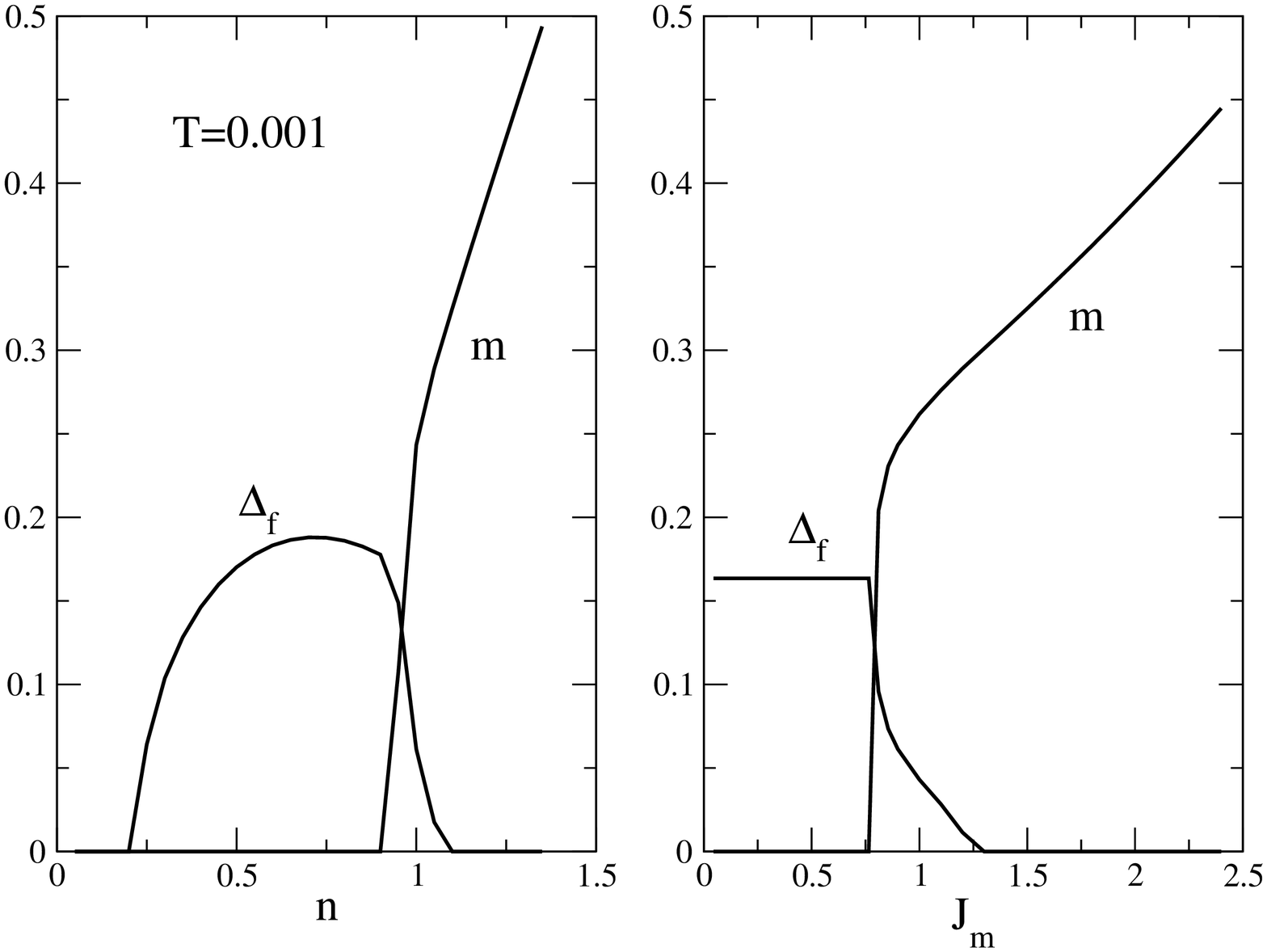}}
\caption{ Magnetization and superconducting order parameters as  functions
of band-filling $n$ in the left panel and of $J_m$ in the right panel
at a low temperature $T=0.001$. 
The other parameters are the same as in Figure 3.
}
\label{mDelvers}
\end{figure}

The phase diagram of the model is quite rich due to the various
correlations and order parameters considered. In this work we focus
our attention on regimes where magnetism and superconductivity
coexist or compete.

In Figure \ref{mDelta} 
we plot the magnetization and the $d$-wave superconducting order parameter
as functions of temperature for a typical case.  The critical
temperature for the antiferromagnetic order parameter $m$,
$T_m$, is larger than the critical temperature, $T_c$, 
for $d$-wave superconductivity. At low temperatures the two phases coexist.
While $\Delta_f \rightarrow 0$ the magnetization $m$ is increasing and then
$m$ decreases to zero at $T_m$ in the usual mean-field like manner.

The left panel of  Figure \ref{mDelvers}
shows the order parameters $m$ and $\Delta_f$ as  functions of
band-filling at a fixed low temperature $T=0.001$. For low to intermediate
band-fillings superconductivity exists. As the band-filling increases, the 
superconductivity is less favorable while the magnetic order parameter appears,
as expected, since the magnetic  order is more favored if the local electron density
is higher. The value of $\Delta_f$ vanishes at low
densities because the $f$-level occupancy also becomes small in that limit
($z \rightarrow 1$, $n_f \rightarrow 0$) and Cooper pairing occurs only
between the $f$-electrons in the model considered.
In the high density limit, $\Delta_f$ also tends
to zero because the  $f$-level occupancy is higher, approaching 1, 
 and freezing of the charge
fluctuations occurs due to the infinite on-site repulsion. Furthermore, a
comparison with  earlier results\cite{peres00}, where magnetic
 order was not considered,
shows that magnetic order lowers the maximum value of the band-filling for
which $\Delta_f \neq 0$, indicating that  the two effects compete with each other.
At smaller band-fillings incommensurate antiferromagnetic order also stabilizes.
Indeed it extends to lower band-fillings as compared to the commensurate case.
However the superconducting order expels the incommensurate case as well and
including both types of ordering the phase diagram is qualitatively similar
for commensurate or incommensurate order if superconductivity is also allowed.

The behavior of  the order parameters against magnetic coupling,
$J_m$, at low temperature, is shown in the right panel of Figure \ref{mDelvers}.
Increasing  $J_m$ leads to a crossover from a region where
$m=0$ to a regime where $m$ grows with $J_m$ while $\Delta_f$ follows the
opposite trend. Keeping $J_m$ fixed and decreasing $J_f$ leads to the opposite
result where the superconductivity disappears in favor of magnetism.

\begin{figure}
\epsfxsize=8.0cm
\epsfysize=8.0cm
\centerline{\epsffile{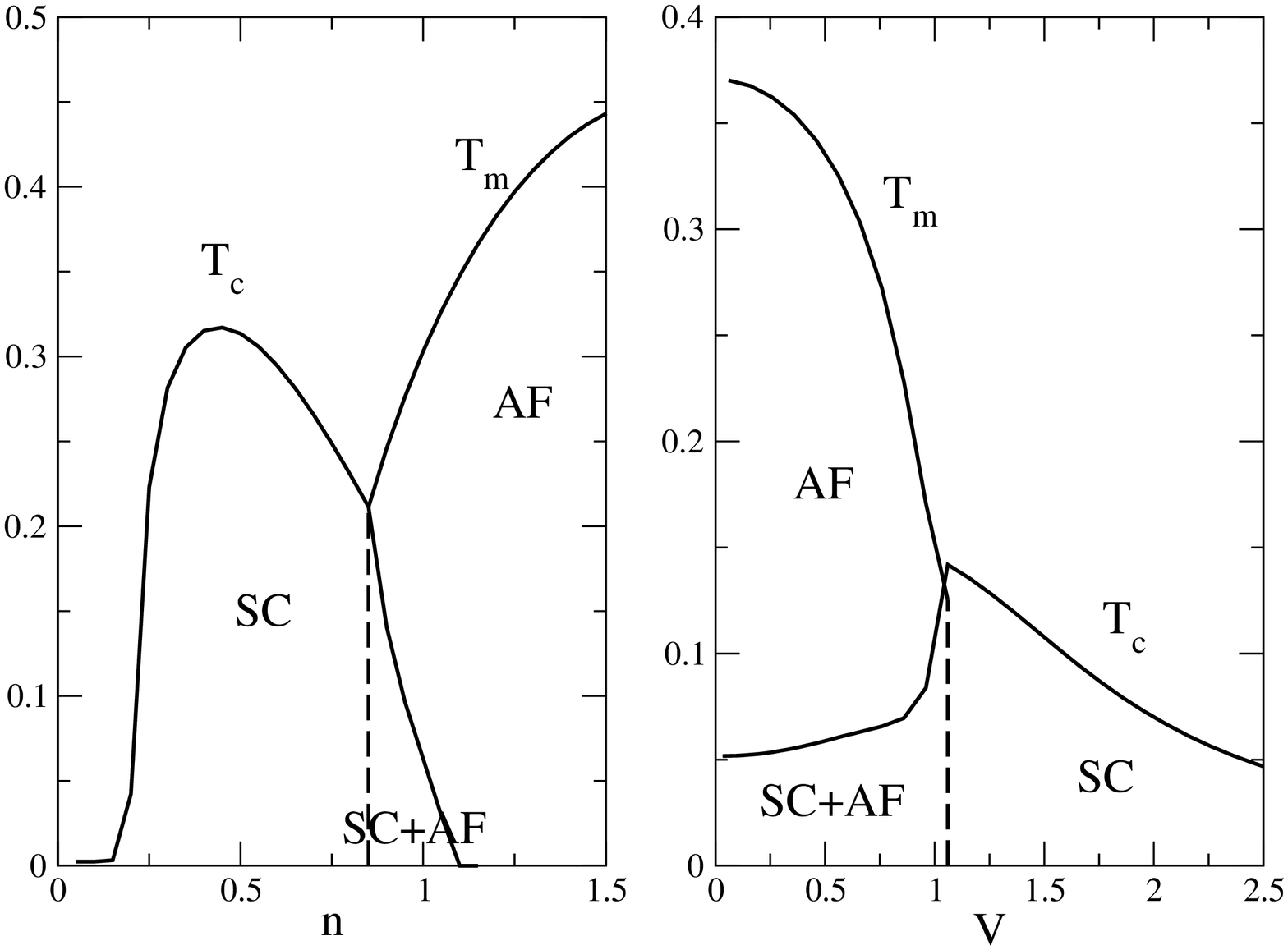}}
\caption{Critical temperatures $T_c$ and $T_m$ as functions of band-filling,
$n$ and of $V$. 
The other parameters are the same as in Figure 3.
The antiferromagnetic (AF), superconducting (SC) and coexisting phases (SC$+$AF) are
shown
}
\label{Tversusnv}
\end{figure}

In Figure \ref{Tversusnv} 
we plot the two critical temperatures as  functions of the
band-filling and of the hybridization.
The behavior of the critical temperatures against density
follows the same trend as in Figure \ref{mDelvers}:
for low $n$ the superconducting 
temperature  is higher while for higher densities  $T_m > T_c$,
indicating which phase is favored as one lowers the temperature from
the disordered high temperature phase. 
Lowering
$\epsilon_0$ or $V$ has the tendency to increase $f$-occupancy favoring
magnetic order  over superconducting order.
At the point where the two temperatures cross 
the magnetic temperature
falls abruptly to zero if the Cooper pairing symmetry is $d$-wave.
This does
not happen if the pairing symmetry is either extended $s$-wave or $p$-wave.
In these cases there is no abrupt expulsion of the magnetization when the
magnetic critical temperature becomes lower than the superconducting critical
temperature and the phase where AF and SC coexist extends for smaller band-filling
values and for larger values of the hybridization.

\begin{figure}
\epsfxsize=8.0cm
\epsfysize=8.0cm
\centerline{\epsffile{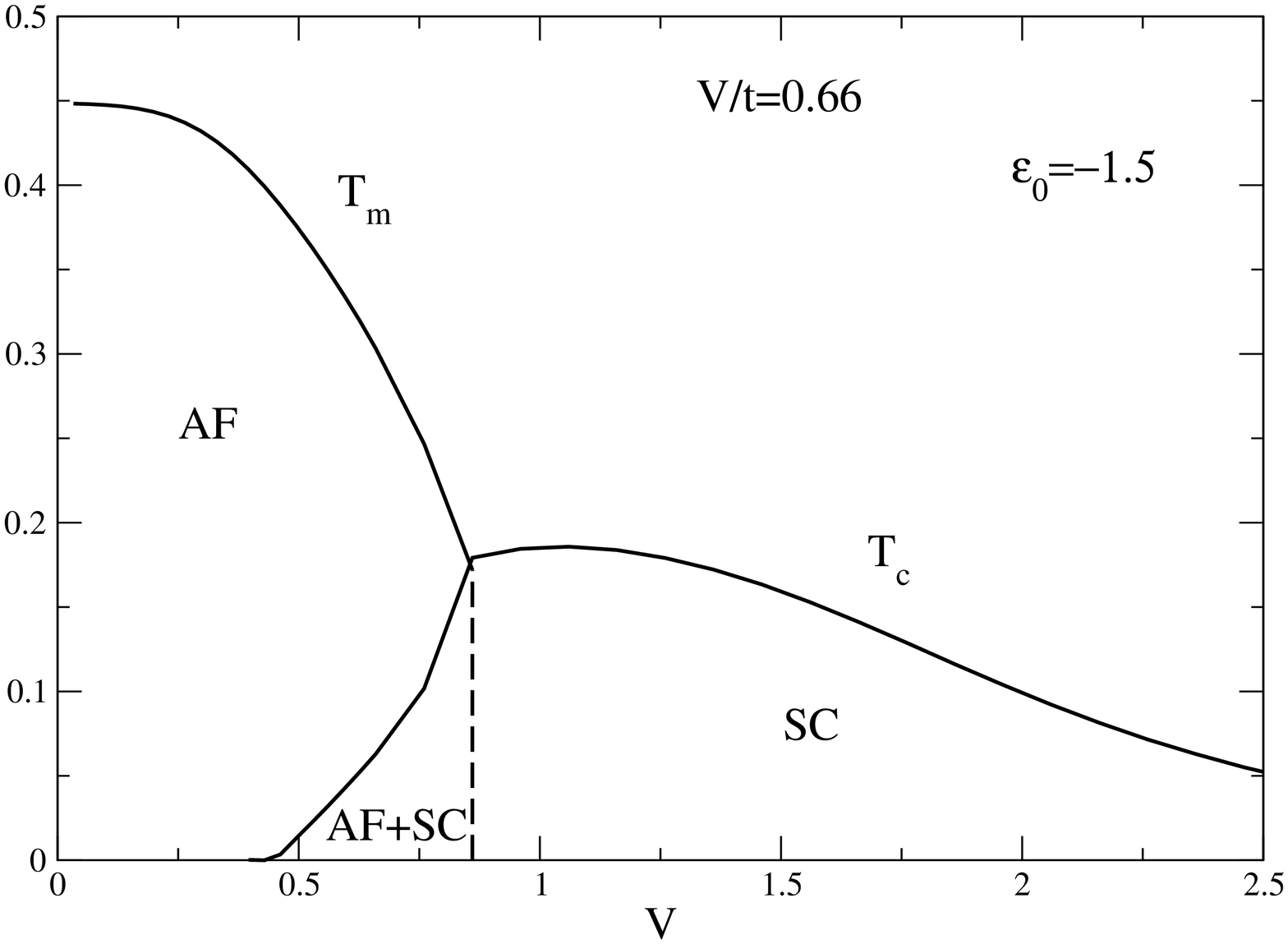}}
\caption{ Critical temperatures $T_c$ and $T_m$ as functions of $V$, with
fixed $V/t=0.66$ for $\epsilon_0=-1.5$.
The other parameters are the same as in Figure 3.
The antiferromagnetic (AF), superconducting (SC) and coexisting phases (SC$+$AF) are
shown
}
\label{Tpressure}
\end{figure}

In the light of the experimental results, it is clearly interesting
to compare qualitatively these results with the experimental phase diagrams, 
where either external pressure or chemical pressure are varied. 
Increasing
pressure is expected to increase both  hopping and  hybridization amplitudes,
while  probably keeping $V/t$ and  other parameters 
approximately constant, to a first approximation\cite{Lacroix}. 
In Figure \ref{Tpressure}
we plot the mean field temperatures as  functions of $V$ for fixed ratio $V/t$.
As pressure increases the magnetic critical temperature decreases
and $T_c$ increases. 
Once again at the point where the two temperatures cross 
the magnetic temperature
falls abruptly to zero if the Cooper pairing symmetry is $d$-wave.
This result, showing expulsion of a spin-density-wave by $d$-wave
superconductivity but not in the case of the other symmetries, is 
very similar to that of a previous study\cite{BR} where the problem of local moment 
formation  in a superconducting phase was addressed: 
increasing pressure  causes expulsion of local moments by $d$-wave 
superconductivity. Therefore, such a phenomenon occurs either in a scenario
of (unscreened or partially screened) ordered local moments or in 
a spin-density-wave scenario, where magnetic order appears as a  Fermi surface
instability. 

The abrupt decrease of $m$ as $T_c$ crosses $T_m$ signals a quantum phase 
transition that can be tuned using the external pressure as a parameter: 
as pressure is reduced (at zero $T$) the groundstate of the system 
changes abruptly 
from nonmagnetic but superconducting to magnetic and superconducting 
at a critical value of $V_c$. The transition appears to be first order.
We note that the same behavior is found for the compound $CeCu_2Si_2$
\cite{Kitaoka}.

\begin{figure}
\epsfxsize=8.0cm
\epsfysize=8.0cm
\centerline{\epsffile{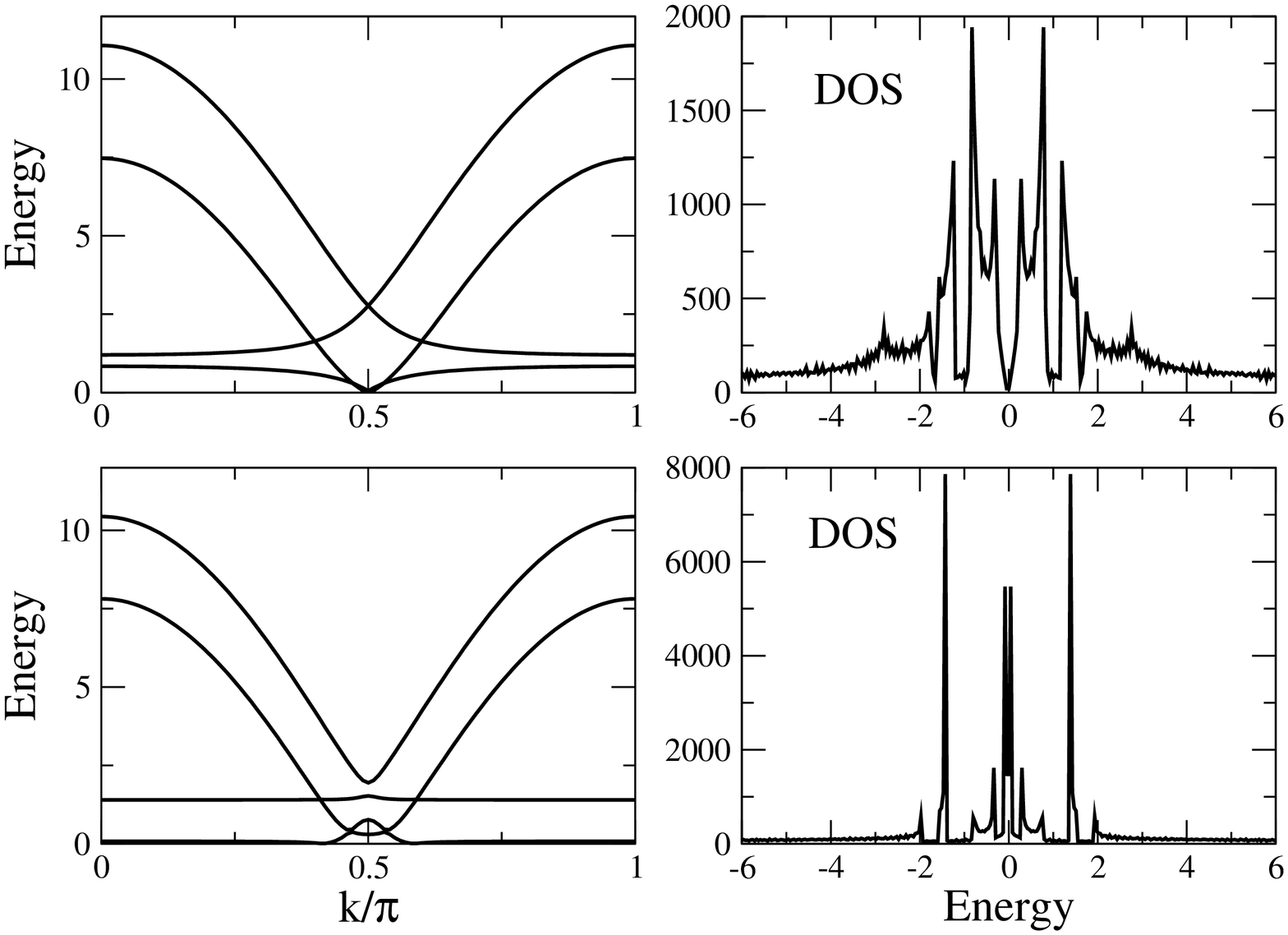}}
\caption{Top panels: Quasiparticle bands and density of states for the set of
parameters $n=1$, $V=1.5$, $t=V/0.66$, $\epsilon_0=-1.5$, $J_f=-3$
and $J_m=0.9$. For this set of parameters the system is in
a phase where $\Delta_f=0.108$, $m=0$ and $z=0.745$.
Lower panels: Quasiparticle bands and density of states for the set of
parameters $n=1.6$, $V=1.5$, $t=V/0.66$, $\epsilon_0=-1.5$, $J_f=-3$
and $J_m=0.9$. For this set of parameters the system is in
a phase where $\Delta_f=0$, $m=0.412$ and $z=0.153$.
}
\label{nomassmass}
\end{figure}

Experimentally, most systems have quite large effective masses. 
Large mass enhancements are obtained when $z$ is small
($n_f \rightarrow 1$). 
This is indeed observed in regimes where the AF
order parameter is large but in these regimes superconductivity is absent.
In Figure \ref{nomassmass} we show results for the quasiparticle bands 
and densities of states where a strong mass enhancement is seen for a large magnetization
in opposition to a situation where the magnetization is zero.
In the calculations above, we have
not been able to find regimes where there is a very large (say larger than $200$)
mass enhancement. 
In the  model considered, the appearance of superconductivity
is restricted to the  mixed
valent regime.  In this situation we do not expect very large densities of 
states at the chemical potential. In a more realistic approach
$U$ is  large but finite and larger band-fillings are allowed
($0 \leq n_f \leq 2$). We expect
therefore that in this regime we might find large effective masses together with
superconducting and magnetic order. 
The restriction to the mixed valent regime is a consequence of the
slave boson approach \cite{Nunes}. In this method the chemical potential is always
pinned to the lowest quasiparticle band, the density is always smaller
than one, and the superconductivity only appears in the
intermediate valence regime with moderate effective masses \cite{IVs}.
Nevertheless, our results  reproduce qualitatively
well the experimental phase diagrams and the competition between the $d$-wave
superconducting and  magnetic phases, such as that observed,
for instance, in $CeCu_2Si_2$. 

\begin{figure}
\epsfxsize=8.0cm
\epsfysize=8.0cm
\centerline{\epsffile{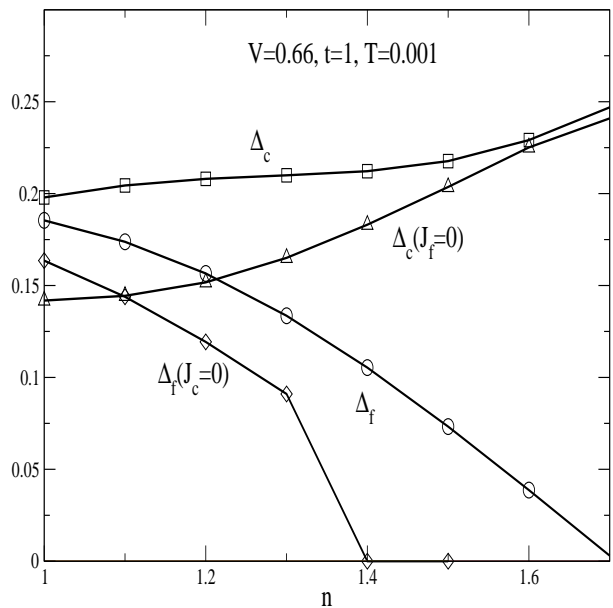}}
\caption{ Superconducting order parameters $\Delta_f$ and $\Delta_c$
(see text) as  functions of band-filling $n$ 
at a low temperature $T=0.001$ where we also include a $c$-electron pairing. 
The other parameters are the same as in Figure 3.
}
\label{cpairing}
\end{figure}

Model (1) is an effective Hamiltonian for the interacting $c$ and $f$ particles. Since
the quasiparticles are heavy close to the top of the lowest band, as evidenced
by the high values of the specific heat jump, we have modelled the superconducting
interaction as taking place between the local $f$ electrons, as usual. One may
wonder however the effect of adding a pairing term, with coupling constant $J_c$,
between the $c$-electrons, since these are present at low energies close to the
chemical potential. Considering a pairing term
in the Hamiltonian in the standard way, with amplitude $\Delta_c$,
we can solve the mean-field equations as before. The effect of this
added term is shown
in Figure \ref{cpairing}. In general, the hybridization exchanges electrons
between the $c$ and the $f$ levels. Due to the restriction on the level
occupancy due to the infinite $U$ repulsion, for a fixed $V$ as the density
increases the number of f-electrons decreases and the superconductivity
is destroyed, as discussed above. However, if the $c$ electrons pair then
the order parameter $\Delta_c$ increases. As a consequence in Figure
\ref{cpairing} we see that as the density increases the ordering in the
c-electrons increases. Actually, if we consider both types of pairing then
the pairing in the f electrons extends to higher densities. So in that
sense adding the pairing between the c electrons favors superconductivity
at higher densities. However, we have found that the adddition of this
type of pairing inhibits the magnetism in the f-electrons probably due to
the greater stability in the pairing channel. 

We remark however, that 
had we written a BCS pairing term only among the $c$-electrons 
in the Hamiltonian (\ref{model}),  
the resulting phase diagram would be
 quite different: the superconducting temperature would be maximum 
at zero hybridization and rapidly decrease with increasing $V$.
This is easily understood because increasing  $V$ makes
$c$-electrons heavier hence reducing their pairing amplitude. 
Clearly this is the opposite trend to that found experimentally in the
phase diagrams as a function of pressure where, at low hybridizations,
superconductivity is absent giving place to the magnetic order.
Even if we do not include the possibility of magnetic order then it
was shown before \cite{peres00} that for small values of the hybridization
the superconducting critical temperature is an increasing function
of pressure.
On the other hand, by having chosen a  pairing interaction between $f$-electrons, 
we obtained increasing $T_c$ since 
the $f$-electrons become more mobile  upon increasing  $V$.
Therefore it is justified to consider only a pairing term between the
heavy f-electrons.

\section{Optical and dynamic conductivity}

The study of the particle-hole excitations of a system can be probed
by studying the finite frequency conductivity. Studies of the Anderson
lattice have been carried out previously \cite{Millis87,Rozenberg}
in the disordered phases and experimental results for the heavy fermions
have been reviewed in Ref \cite{deGiorgi}.
 In traditional superconductors the optical
conductivity clearly shows a threshold at twice the 
gap energy\cite{Bickers,Chen}. Since the
d.c. conductivity of a perfect lattice is infinite, most studies are
carried out taking into account scattering off impurities \cite{Mattis}
considering moderate scattering up to the dirty limit \cite{Chen,Mattis}.
In this work we will study the optical conductivity in the clean limit
and we shall
study its behavior in the ordered phases as well.

\begin{figure}
\epsfxsize=8.0cm
\epsfysize=8.0cm
\centerline{\epsffile{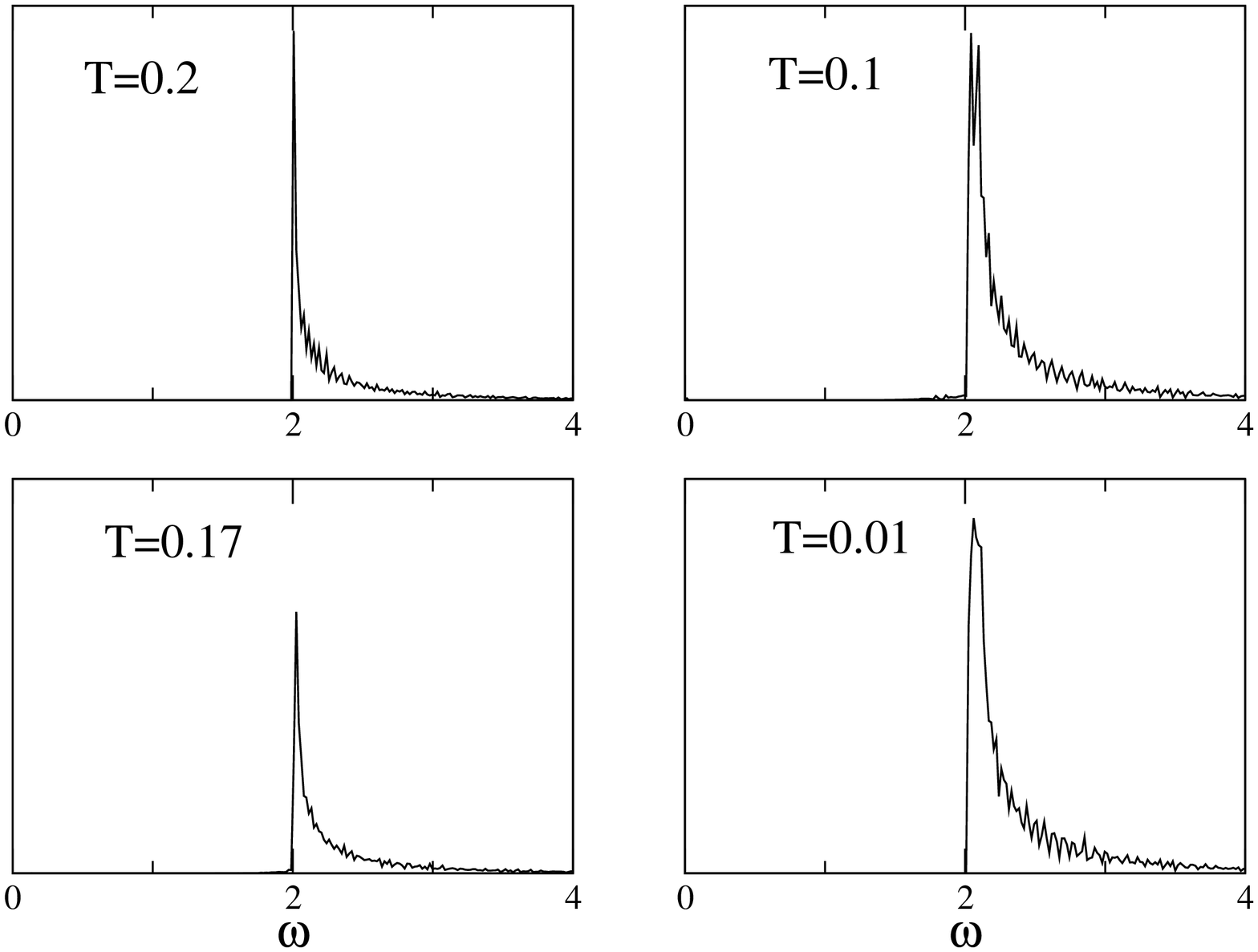}}
\caption{ Optical conductivity (in arbitrary units) 
at several temperatures. $n=1$, $V=1.2$, $t=V/0.66$, $\epsilon_0=-1.5$,
$J_f=-3$, $J_m=0.9$. 
For these parameters
the system is superconducting at low temperatures and has no magnetization.
}
\label{sig1}
\end{figure}

\begin{figure}
\epsfxsize=8.0cm
\epsfysize=8.0cm
\centerline{\epsffile{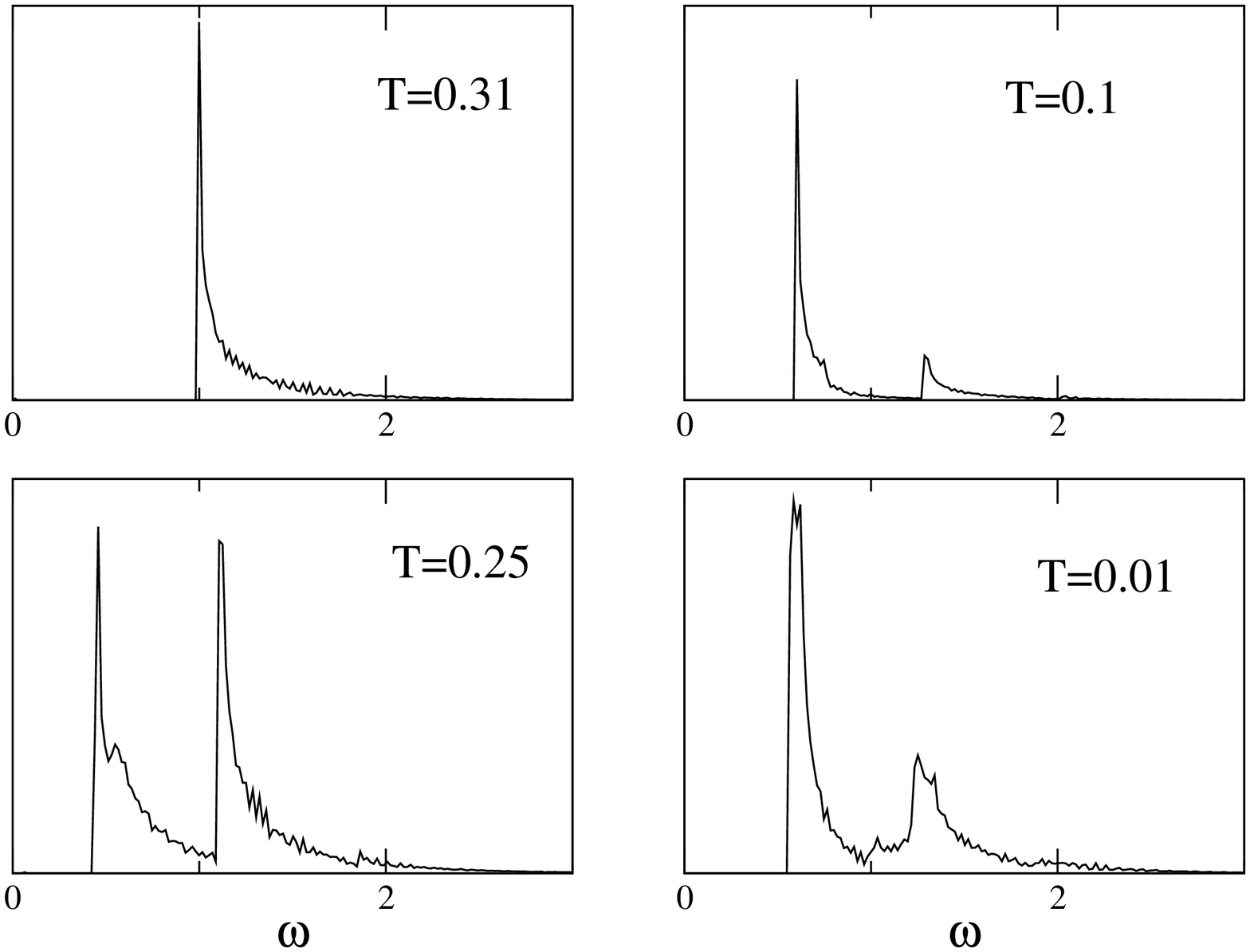}}
\caption{Optical conductivity (in arbitrary units) at several temperatures. 
$n=1$, $V=0.7$, $t=V/0.66$, $\epsilon_0=-1.5$,
$J_f=-3$, $J_m=0.9$.  
For the parameters chosen 
magnetism and superconductivity coexist at low temperatures.
}
\label{sig2}
\end{figure}

The real part of the dynamic conductivity is given by
\be
\sigma_{\alpha \beta} (\vec{q},\omega)  = \frac{1-e^{-\beta \omega}}{2 \omega V}
\int_{-\infty}^{\infty} dt e^{i \omega t} <j_{\alpha}^{\dagger} (\vec{q},t)
j_{\beta}(\vec{q},0)>
\ee
Writing the trace, inserting a decomposition of the identity in terms
of the exact many-body energy states and integrating over time we obtain
\bea
\sigma_{\alpha \beta} (\vec{q},\omega)  & = & \frac{1-e^{-\beta \omega}}{2 \omega V}
2 \pi \sum_{n,m} \frac{e^{-\beta E_n}}{Z} \delta (\omega +E_n-E_m) \nonumber \\
& & <n|j_{\alpha}^{\dagger} (\vec{q}) |m><m|j_{\beta} (\vec{q})|n>
\eea
Consider now $\alpha=\beta$ and $\vec{q}=0$. 
The $\alpha=x,y$ component of the current operator can be written as
\be
j_{\alpha} = -it \sum_{\vec{k},\sigma} \left( \vec{\chi}_{\vec{k}} \right)_{\alpha} 
c_{\vec{k} \sigma}^{\dagger}
c_{\vec{k}\sigma}
\ee
where $\vec{\chi}_{\vec{k}} = \sum_{\vec{\delta}} \vec{\delta} 
e^{-i \vec{k} \cdot \vec{\delta}}$
where $\sigma$ is the spin component. Writing the electronic
operators in terms of the Bogolubov operators it is
straightforward to calculate the matrix elements of the
current operator. 

We can also calculate the finite-momentum and finite-frequency
conductivity $\sigma_{xx}(\vec{q},\omega)$. Starting from eq. (8)
and taking similar steps to those followed for the calculation of the
optical conductivity it is easy to calculate the dynamical
conductivity.

In Figures \ref{sig1} and \ref{sig2}
 we show the optical conductivity  in the clean
limit for two typical cases. Figure  \ref{sig1} refers to a regime where there is
superconductivity but no magnetization at low temperature.
Figure  \ref{sig2} refers to 
a regime where there is  coexistence of superconductivity
and antiferromagnetism  at low temperature.

The experimental study of the optical conductivity,
over a large range of frequencies,
of some heavy fermion systems has only recently become available.
\cite{dordevic,dressel}
In reference \cite{dressel} it was found that effective mass $m^\ast$
of the quasiparticles increased about 50 times when the temperature
decreases below the N\'eel temperature. Furthermore, the authors found
for $T<T_N$ that a two peak structure developed at finite frequencies.
At temperatures higher than
$T_N$ the $\sigma(\omega)$ presents a single peak, separated from
the Drude weight by a finite gap (if disorder is included the
Drude peak broadens to finite energies). When the temperature
drops below $T_N$ a second peak, respecting to
the magnetic gap, shows up in $\sigma(\omega)$ at lower, but finite, energies. 
The features reproduced by our calculation are in qualitative
agreement with the data presented in Ref. \cite{dressel}.
Also they are in qualitative agreement with the results of \cite{Sato}.
Note however that our model is not appropriate to describe this material
since one needs to take into account the dual nature of the f-electrons,
which is not included in our model. This shows that the results obtained
are qualitative general trends that are captured by our simplified model.

\begin{figure}
\epsfxsize=8.0cm
\epsfysize=8.0cm
\centerline{\epsffile{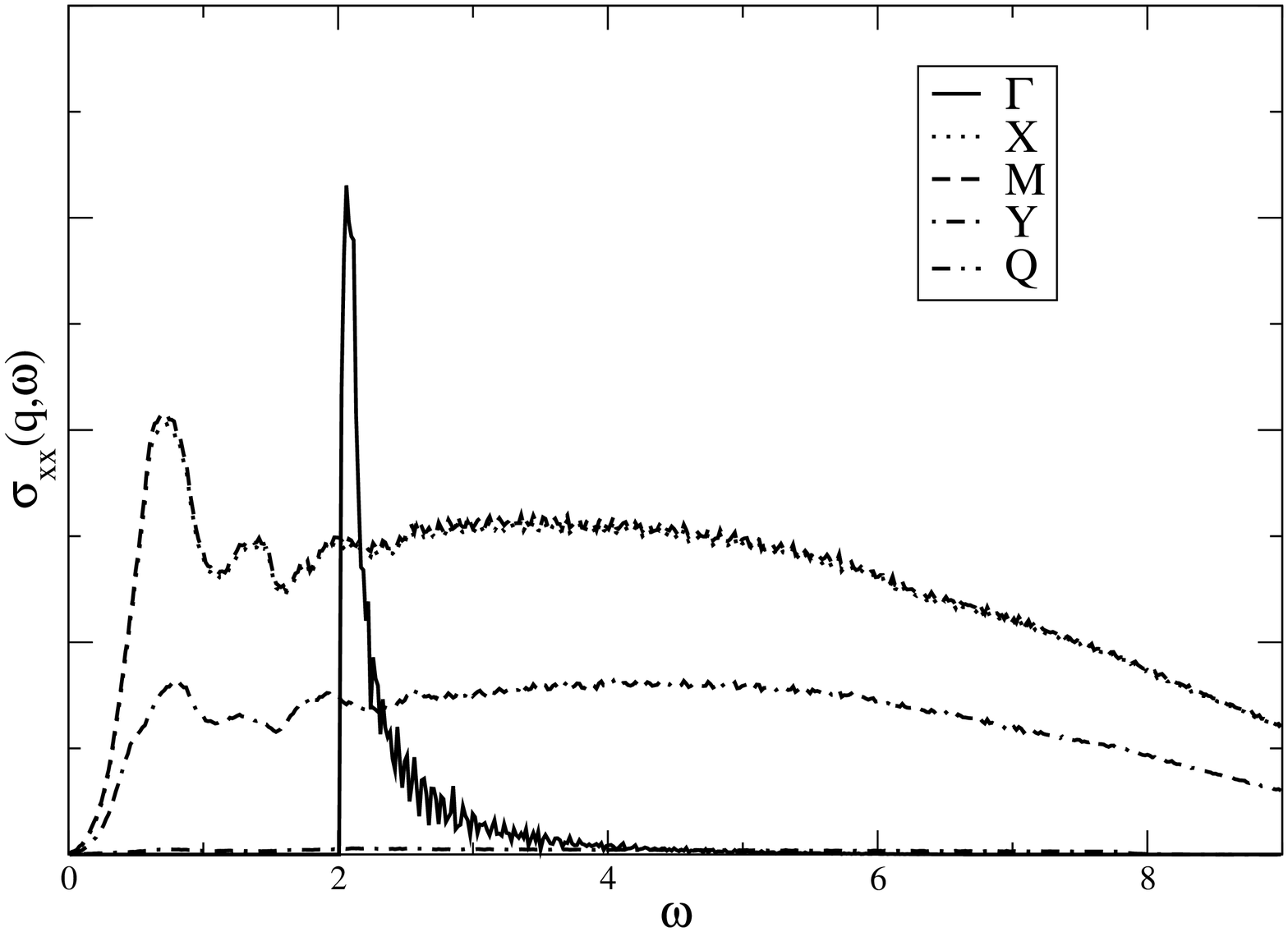}}
\caption{ Finite-momentum and finite frequency conductivity at several
points in the Brillouin zone for a set of parameters such that the system
is superconducting ($\Gamma=(0,0), X=(\pi,0), M=(\pi,\pi), Y=(0,\pi), Q=(\pi/2,\pi/2)$). 
The temperature is $T=0.01$ and $V=1.2$. The results
for the $\Gamma$ point, $\vec{q}=0$, are multiplied by 1000.
}
\label{mom1}
\end{figure}

\begin{figure}
\epsfxsize=8.0cm
\epsfysize=8.0cm
\centerline{\epsffile{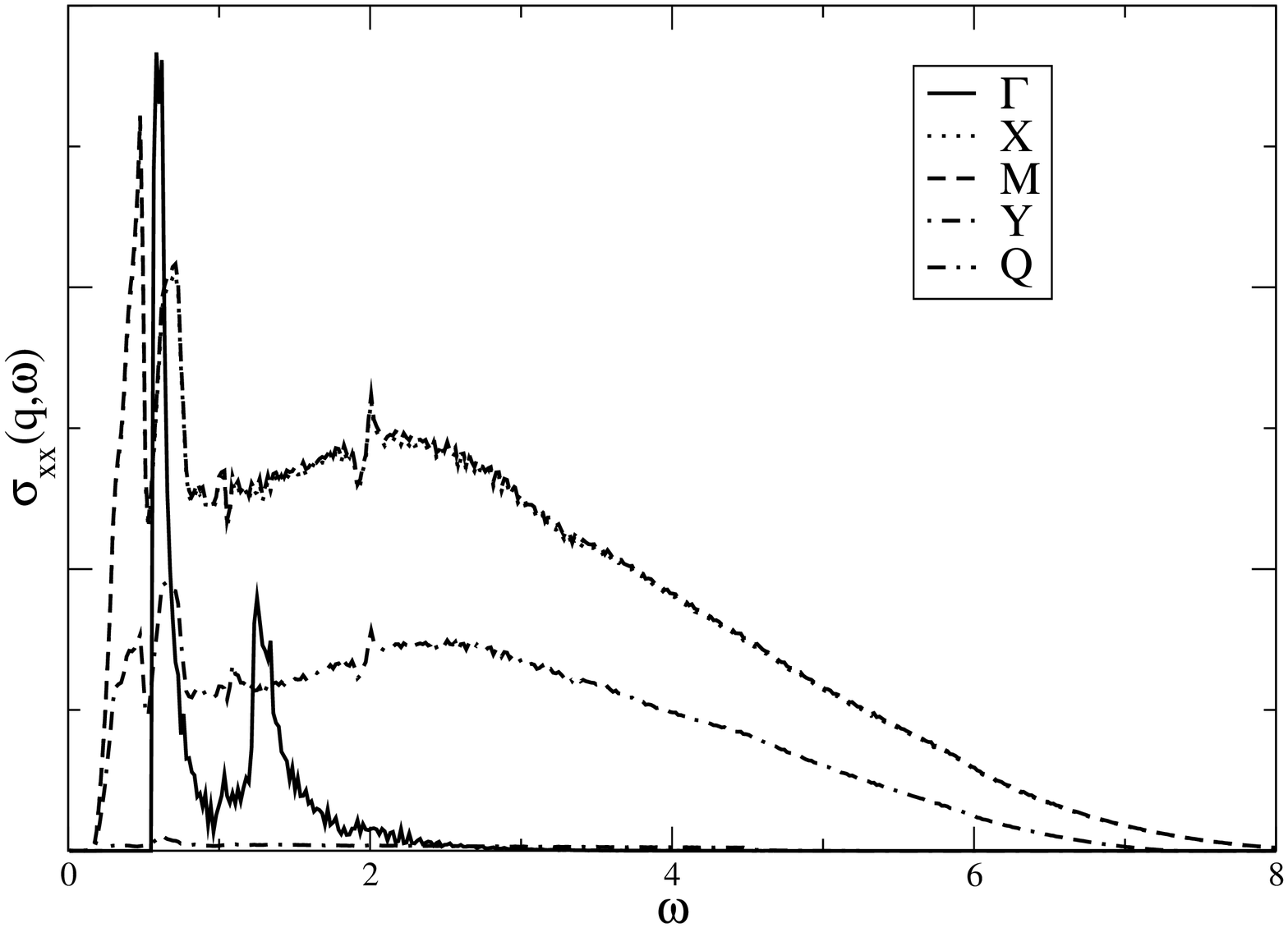}}
\caption{Finite-momentum and finite frequency conductivity at several
points in the Brillouin zone for a set of parameters such that the system
is superconducting and magnetic. The temperature is $T=0.01$ and $V=0.7$.
The results
for the $\Gamma$ point, $\vec{q}=0$, are multiplied by 1000.}
\label{mom2}
\end{figure}

\begin{figure}
\epsfxsize=8.0cm
\epsfysize=8.0cm
\centerline{\epsffile{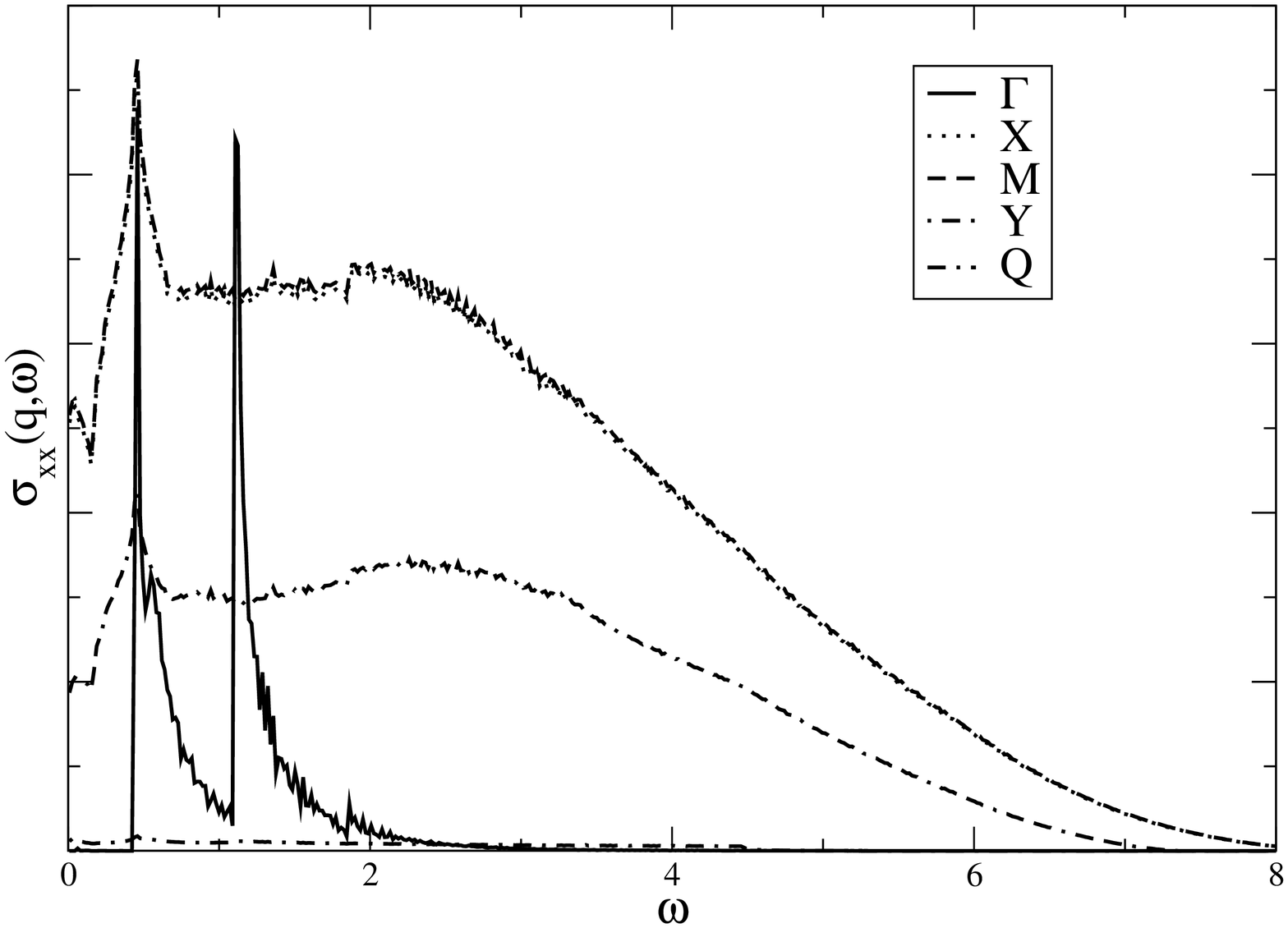}}
\caption{Finite-momentum and finite frequency conductivity at several
points in the Brillouin zone for a set of parameters such that the system
is magnetic. The temperature is $T=0.25$ and $V=0.7$.
The results
for the $\Gamma$ point, $\vec{q}=0$, are multiplied by 1000.}
\label{mom3}
\end{figure}

In order to preserve momentum the optical conductivity probes
the transitions between different bands coupled by the current
operator. If we allow that momentum is interchanged then the
dynamic conductivity also probes excitations along the same
band in addition to across bands.
In Figs. \ref{mom1}, \ref{mom2}, \ref{mom3} we present the dynamic conductivity 
for different points in
the Brillouin zone in three typical situations. In Fig. \ref{mom1} the set of parameters
and temperature are such that the system is superconducting, in Fig. \ref{mom2} the
system is both superconducting and magnetic and in Fig. \ref{mom3} the system is only
magnetic. The gap structure evident in the optical conductivity ($\vec{q}=0$) is now absent
because
 low energy excitations are in general allowed since there is coupling
between states in the same band. In the superconducting case (but not magnetic)
the dynamic conductivity extends to zero frequency but vanishes in this limit
due to the energy dispersion of the $d$-wave symmetry. If there is
coexistence of superconductivity and magnetism then the same structure appears
with the double-peak feature, characteristic of the magnetic phase, enhanced,
as observed in the optical conductivity. If the superconducting order parameter
is zero then the dynamic conductivity is finite at zero energy.
 
\section{Summary}

The interplay between magnetic correlations, the Kondo effect
and superconducting correlations in heavy fermion systems
is a difficult problem to solve. While previous studies on the 
$U=\infty$
Anderson lattice have identified the instabilities towards magnetic
or superconducting order by taking into account the slave-boson fluctuations,
 the description of the ordered phases has not previously been carried out.
 In this work we have phenomenologically
studied the interplay between the superconducting and antiferromagnetic
ordered phases by studying their dependence on  the model parameters.
 We have found that Cooper pairing symmetry plays an important role in
determining the regimes of either coexistence or competition between 
superconductivity and antiferromagnetism: indeed, we have found that 
$d$-wave superconductivity  coexists with antiferromagnetic order
but the former expels the latter abruptly as the hybridization 
between the $f$-level and the conduction electrons increases. We also found
that the $f$-level occupancy affects the two types of order in 
different ways: higher occupancy favors antiferromagnetism while a
lower occupancy favors superconductivity. Hence, 
the quasiparticle mass enhancement relative to the bare electron
was found not to exceed a few tens in the superconducting phase. The optical conductivity 
has also been computed,  reflecting the band structure of
the ordered phases.

\ack

The author wants to thank Miguel Ara\'ujo and Nuno Peres for many
discussions on this subject. 

\end{document}